\documentstyle[12pt,iopfts]{ioplppt}

\begin{document}

\title{Toroidal quadrupole transitions associated to
collective rotational-vibrational motions of the nucleus}
[Toroidal quadrupole transitions]
\author{\c S Mi\c sicu}
\address{ National Institute for Earth-Physics,
Bucharest Magurele, P.O.Box Mg-2, ROMANIA}

\begin{abstract}
In the frame of the collective model one computes the
longitudinal, transverse and toroidal multipoles corresponding to the
excitations of low-lying levels in the ground state band of several
even-even nuclei by inelastic electron scattering $(e,e')$. In connection
with this transitions a new quantity, which accounts for the
deviations from the Siegert theorem, is introduced in the frame of the
Riemann Rotational Model with non-vanishing vorticity.
Inelastic differential cross-sections calculated at backscattering
angles shows the dominancy of toroidal form-factors over a broad range of
momentum transfer
\end{abstract}

\section{Introduction}

A still opened question in nuclear structure theory concerns the
nature of current flows inside the nucleus. Many attempts have been made up
to the present time in order to clarify whether the nuclear matter can be
portrayed as a quantum fluid that could support only irrotational flows
(IF) or is a quantum rotator which give rise to rigid rotation of the whole
nucleus (RR) [1]. There have been also proposed intermediate models, like
the Riemann rotational model [2] , in which the velocity field is assumed to
be a linear combination of rigid and irrotational contributions.
Since these models give quite different descriptions for the
collective current distributions, a direct determination of the nuclear
current is required. The measurement of transverse multipoles in inelastic
electron scattering, provides a practical tool for this purpose
[3]. The longitudinal (Coulomb) multipole is related to the charge
distribution of the nucleus due to the charge-current conservation law. In
order to obtain the quantities which depend on the nuclear current one have
to measure the transverse part of the cross-section, i.e. the electric and
magnetic multipoles.

There is a large amount of information concerning the theoretical
calculations and experimental determination of electromagnetic
multipoles for such reactions. Our aim is to qualitatively investigate the
transverse electric multipoles from the point of view of a multipole
decomposition introduced by Dubovik and Cheshkov [4]. This decomposition,
contains in the second order of momentum transfer the toroidal multipole
moment which is independent of the other two classes of known
electromagnetic multipoles: the charge and the magnetic one. The first
member of this new class of electromagnetic multipoles is called the $
toroidal~dipole$ or $anapole$ moment, and was introduced by Zel'dovich [5]
in connection with the parity violation in weak interactions. The second
member is the $toroidal~quadrupole$ or $tetrapole$ which is a parity-even
and time-odd electromagnetic characteristic [6-8].
In this paper we are interested in the dynamic counterpart, i.e.
the induced toroidal moments. Such characteristics turn out to be associated
with a isoscalar $\lambda ^\pi =1^{-}$ state, called the dipole
torus mode, recently predicted in the frame of nuclear fluid
dynamics [9,10].
Calculations of the electromagnetic form factors entering into
the photoabsorbtion or electron scattering cross sections show that
only the toroidal form factor contributes to the excitation of this mode
[11]. Therefore for such excitation in the low-$q$ limit we obtain the
toroidal dipole transition, not the charge dipole one as is stated by the
Siegert theorem for the case of electric GDR.\\ \\

\section{Toroidal multipoles}

The study of normal-parity excitations in nuclei require the
knowledge of longitudinal(Coulomb) and transverse electric multipoles. For
simplicity we consider their expressions in Born approximation (PWBA)
\begin{equation}
\hat T_{\lambda \mu }^L=\frac{i^{\lambda +1}}{\sqrt{2\lambda
+1}}(\sqrt{\lambda }\delta _{\lambda ^{\prime },\lambda -1}+\sqrt{\lambda
+1}\delta_{\lambda ^{\prime },\lambda +1})
\int {d^3rj_{\lambda ^{\prime}}(qr)
{\bi Y_{\lambda \lambda ^{\prime }}^\mu (\Omega )}
\cdot {\hat\bi J}({\bi r})}
\end{equation}
\begin{equation}
\hat T_{\lambda \mu }^E=\frac{i^{\lambda +1}}{\sqrt{2\lambda+1}}
(\sqrt{\lambda +1}\delta _{\lambda ^{\prime },\lambda -1}-\sqrt{\lambda}
\delta_{\lambda ^{\prime },\lambda +1})\int {d^3rj_{\lambda ^{\prime}}(qr)
{\bi Y_{\lambda \lambda ^{\prime }}^\mu (\Omega )}
\cdot {\hat\bi J}({\bi r})}
\end{equation}
where
\begin{equation}
{\bi Y_{\lambda \lambda ^{\prime }}^\mu (\Omega )}=[Y_{\lambda^{\prime}}
(\Omega )\otimes {\bi e}_1]_\lambda ^\mu
\end{equation}
are the vector spherical harmonics [12] and ${\hat\bi J}({\bi r})$ is the
proton current. Choosing the z-direction to be along the momentum
transferred from the electron to the nucleus ${\bi q}$, the longitudinal and
transverse electric reduced matrix elements (r.m.e.) are
\begin{equation}
t_{L\lambda }(q)_{fi}\equiv \langle I_f\mid \mid {\hat T_{\lambda}^L}(q)
\mid \mid I_i\rangle
\end{equation}
\begin{equation}
t_{E\lambda }(q)_{fi}\equiv \langle I_f\mid \mid {\hat
T_{\lambda}^E}(q)\mid \mid I_i\rangle
\end{equation}
For low momentum transfer, one sees from eq.(1) that Siegert's theorem [13]
\begin{equation}
t_{E\lambda }(q\rightarrow 0)_{fi}=
\sqrt{\frac{\lambda +1}\lambda }t_{L\lambda }(q\rightarrow 0)_{fi}
\end{equation}
is satisfied, independently of the approximation used for the nuclear
current. While this relation is very important for low-energy real-photon
processes, it is not suitable at higher momentum transfer.
Following the ideas outlined in [4,14], one can build a variant of Siegert's
theorem at arbitrary momentum $q$.
Thus the r.m.e. of the transverse electric multipole $t_{E\lambda}(q)$
can be separated in two parts: one depending on the low momentum transfer
$q$ limit and the other on the high-$q$ terms
\begin{equation}
t_{E\lambda }(q)_{fi}=\sqrt{\frac{\lambda +1}\lambda}t_{L\lambda}(0)_{fi}+
{q^{2}}t_{T\lambda }(q)_{fi}
\end{equation}
where $t_{T\lambda }(q)_{fi}$ is the r.m.e. of the toroidal multipole
operator
\begin{equation}
\hat T_{\lambda \mu }^T=\frac{i^\lambda }{q^{3}}\int
{d^3r{\bnabla }\times
{\Large\{}[j_\lambda (qr)-\frac{(qr)^\lambda }{(2\lambda +1)!!}]
{\bi Y_{\lambda\lambda }^\mu (\Omega )}{\Large\}}\cdot {\hat\bi
J}({\bi r})}
\end{equation}
Therefore $t_{T\lambda }(q)$ contains the whole information of
$t_{E\lambda}(q)$ for high-$q$. In the longwavelength limit the toroidal
multipole $t_{T\lambda }(0)$ is proportional to the transition toroidal
multipole moment ${\bcal T}_{\lambda}$
\begin{equation}
t_{T\lambda }(0)=-\frac{(iq)^{\lambda-1}}{(2\lambda
+1)!!}\sqrt{\frac{
\lambda +1}\lambda }{\bcal T}_{\lambda }
\end{equation}
For ${\lambda =1}$, we have the earlier mentioned toroidal dipole moment
vector operator
\begin{equation}
{\hat{\bf T}}=\frac{1}{10}\int {d^3r[{\bi r}({\bi r}\cdot
{\hat\bi J}({\bi r}))-2r^2{\hat\bi J}({\bi r})]}
\end{equation}
This electromagnetic moment is associated with a poloidal current flow on
the wings of a toroidal solenoid. More details can be found in the excellent
review [14]. The next terms in the $q$-expansion of the toroidal multipoles
are the average 2n-power radii of the current distribution
$\rho_{\lambda \mu}^{2n}$, which give rise to the corresponding toroidal
multipole moment. Their general form is
\begin{equation}
\hat\rho_{\lambda\mu}^{2n}=\int{d^{3}r~r^{\lambda+2n+1}}
({{\bi Y_{\lambda \lambda-1}^\mu
(\Omega)}+\frac{2n+2}{2\lambda+2n+1}
\sqrt{\frac{\lambda}{\lambda+1}}{\bi Y_{\lambda \lambda+1}^\mu
(\Omega )}})\cdot{\hat\bi J}({\bi r})
\end{equation}
We end this section by noticing that from the point of view of
other multipole expansion treatments, the toroidal moments can be
viewed as electric multipoles induced by magnetization [15] or as
retardation corrections to the Siegert theorem [16].

\section{Deviations from Siegert's theorem in the collective model}

In order to illustrate the ideas presented in the previous chapter we
consider the Riemann Rotational Model, which is an algebraic extension
of the Bohr-Mottelson model. This model assumes that the velocity
field is a linear function of the position. A Riemann rotor is characterized
by a parameter ${\tilde r}$, called rigidity, which ranges from zero to one.
If ${\tilde r}$=1 the body is rotating rigidly (RR) with the angular
frequency $\bomega$; when ${\tilde r}$=0 the flow is irrotational (IF) and
the nuclear surface performs harmonic oscillations with amplitudes given
by the collective coordinates $\alpha_{lm}$.
The rigidity is related to the quantity called vorticity ${\bcal L}$
[17,18]. The velocity vector field of a Riemann rotor is a convex
combination of the rigid and irrotational values,
\begin{equation}
{\hat\bi J(\bi r)}={\tilde r}{\hat\bi J_{RR}(\bi r)}+
(1-{\tilde r}){\hat\bi J_{IF}(\bi r)}
\end{equation}
where the RR and IF velocity fields are given by
\begin{equation}
{\bi v}_{RR} = {\bi \bomega}\times{\bi r}
\end{equation}
and
\begin{equation}
{\bi v}_{IF} = {R^{2}}\sum_{lm}\frac{\dot \alpha_{lm}}{l}
[{\bnabla}({r\over R})^l]Y_{lm}(\theta,\phi)]\theta(R-r)
\end{equation}
Introducing the above equations in the expressions of longitudinal and
transverse electric multipoles, for an axial-symmetric nucleus
with quadrupole deformation $\beta$ one get
\begin{equation}
\hat T_{2 \mu }^{L}(q,{\tilde r}) =
-Ze\sqrt{\frac{2}{\pi}}\frac{\sqrt{30}}{40}
\frac{Q_{0}}{R}\mu\omega_{\mu}(j_{1}(qR)+j_{3}(qR))
\end{equation}
and
\begin{equation}
\hat T_{2 \mu }^{E}(q,{\tilde r}) =
-Ze\sqrt{\frac{3}{\pi}}\frac{\sqrt{30}}{40}
\frac{Q_{0}}{R}\mu\omega_{\mu}(j_{1}(qR)+
(1-\frac{5}{3}{\tilde r})j_{3}(qR))
\end{equation}
Here $-Ze$ is the nuclear charge, $Q_{0}=\sqrt{3\over5\pi}R^{2}\beta$ is
the static quadrupole moment and $R=r_{0}A^{1/3}$.
One thus observe that the longitudinal multipoles are insensitive
to the rotational components of the velocity field, since their value is
constant for any value of the rigidity parameter ${\tilde r}$.
Moreover this value is proportional to the transverse multipole for
irrotational flow
\begin{equation}
\hat T_{2 \mu}^{L}(q,{\tilde r})
= \sqrt{\frac{2}{3}}\hat T_{2\mu}^{E}(q,{\tilde r}=0)
\end{equation}
The next step consists in the study of low momentum transfer
behaviour of the above discussed multipoles. One obtain the well known
$Siegert~theorem$ which states that in the long-wavelength limit the
transverse electric multipole can be expressed in terms of the longitudinal
one
\begin{equation}
\hat T_{2 \mu }^{E}(q\approx 0,{\tilde r}) = \sqrt{\frac{3}{2}}
\hat T_{2 \mu }^{L}(q\approx 0,{\tilde r})
\end{equation}
Recalling the remark that we have made above about the constancy of the
longitudinal multipole with ${\tilde r}$, one conclude that the
proportionality between this multipole and the transverse electric one is
satisfied for the IF for arbitrary momentum transfer $q$. For the RR this
relation is valid only for low-$q$ transfer. This means that in principle
the reactions performed at low-$q$ are not able to give informations about
the vorticity. The Riemann rotator behaves as an irrotational liquid drop
at small transferred momentum. From the point of view of the multipole
expansion this fact can be understood invoking the following argument : in
the low-$q$ limit the longitudinal multipole is proportional to the time
derivative of the charge quadrupole moment operator
\begin{equation}
\hat T_{2\mu}^{L}(q\approx 0,{\tilde r})
= -\frac{iq}{15}{\dot {\hat Q}}_{2\mu}
\end{equation}
where the charge quadrupole moment has the well-known form
\begin{equation}
\hat Q_{\lambda\mu} = \int d^{3}r~r^{\lambda}Y_{\lambda\mu}
\hat \rho({\bi r},t)
\end{equation}
According to the charge-current conservation law
\begin{equation}
\frac{\partial \hat\rho}{\partial t} + {\bnabla}\cdot{\hat\bi J} = 0
\end{equation}
the time derivative of the charge quadrupole ${\dot {\hat Q}}_{2\mu}$
may be put in the form
\begin{equation}
{\dot {\hat Q}}_{2\mu} = \int d^{3}r~r^{\lambda}Y_{\lambda\mu}
{\bnabla}\cdot{\hat\bi J}
\end{equation}
Since the gradient cancels the rotational (vortical)
components of the nuclear current, the time derivative of the
charge quadrupole moment is to be associated with curless
quadrupole flows, i.e. $\bnabla\times{\hat\bi J}=0$,
like the ${\beta}$ and ${\gamma}$ vibrations. This is the reason why the
nuclear response is vibrating-like (without shear components) for low-$q$
transfer. Therefore in order to obtain informations about the rotational
currents inside the nucleus one must go beyond the Siegert theorem by
increasing the energy transferred in the scattering reaction. The high-$q$
terms arising in the multipole expansion of the transverse electric
multipole are free of the charge-current conservation constraint. Hence
they could provide us with information about the vortical components, i.e.
$\bnabla\times{\hat\bi J}\neq 0$. In order to accomplish this task we
employ the above mentioned multipole expansion formalism for the
transverse electric multipoles, where the Siegert limit term is
explicitly decoupled from the high-$q$ terms. Equation {(7)} is
rewritten in the quadrupole case as follows by factorizing the
longitudinal multipole in the $q=0$ point
\begin{equation}
t_{E2}(q,{\tilde r})_{fi}={\sqrt\frac{3}{2}}t_{L2}(0)_{fi}
(1-\eta_{2}(q,{\tilde r}))
\end{equation}
In the above equation we have introduced a quantity related to
the deviation from the Siegert theorem that is also given as a
convex combination of rigid and irrotational values
\begin{equation}
\eta_{2}(q,{\tilde r})\equiv
\frac{q^{2}t_{T2}(q,{\tilde r})_{fi}}{t_{E2}(0,{\tilde r})_{fi}}
={\tilde r}~{\eta}_{2}(q,{\tilde r}=1)+
(1-{\tilde r})~{\eta}_{2}(q,{\tilde r}=0)
\end{equation}
In Fig.1 a plot of this quantity vs. momentum transfer is
presented. The RR model displays a stronger deviation from the Siegert
theorem than the IF model. For the hexadecupole case ($\lambda=4)$, the
effect is lowered with respect to the quadrupole case ($\lambda=2)$ for both
models. Another quantity of interest defined in electron scattering
theory [19] is the real transverse electric form factor
\begin{equation}
F_{\lambda}^{E}(q)\equiv\frac{t_{E\lambda}(q)}{t_{E\lambda}(0)}
\end{equation}
with ${F_{\lambda}^{E}(0)=1}$. Specializing to the quadrupole case
of the Riemann Rotator, expanding this form factor according to the
Dubovik-Cheshkov procedure and using the definition of the toroidal
multipole moment [eqs.(9) and (19)] one gets in the first order of $q^{2}$
\begin{equation}
F_{2}^{E}(q)\approx 1 - \frac{{q^2}}{3}{{\bcal I}_{{\tilde r}}}\cdot
\frac{{\bcal T}_{2}}{Q_{2}}
\end{equation}
where ${\bcal I}_{{\tilde r}}$ is the inertia moment of the Riemann
Rotator and $Q_{2}=3e^{2}ZR_{0}^{2}\beta/4\pi$ is the r.m.e. of the
transition charge quadrupole moment, which does not depend on the rigidity
parameter ${\tilde r}$.
\begin{table}
\caption{Toroidal quadrupole transition moment ${\bcal T}_{2}$ in
arbitrary units at fixed deformation $\beta$ for $^{152}$Sm, $^{154}$Sm
and $^{166}$Er for three different values of the rigidity
parameter ${\tilde r}$ :
IF (${\tilde r}=0$), RR (${\tilde r}=1$) and for the Riemann Rotational
Model
with ${\tilde r}=0.5$.}
\begin{indented}
\item[]\begin{tabular}{@{}llll}
\br
Nucleus &~~~~~~~~~
$\beta$~~~~~~~~~&~~~~~~~~~${\tilde r}$~~~~~~~~~&~~~~~~~~~${\bcal
T}_{2}~~~~~~~~~$\\
\br
 & & 0. & 5.2$\cdot$10$^{-4}$\\
$^{152}$Sm & 0.246 & 0.5 & 1.32$\cdot$10$^{-3}$\\
 & & 1. & 1.72$\cdot$10$^{-2}$\\
\mr
 & & 0. & 5.6$\cdot$10$^{-4}$\\
$^{154}$Sm & 0.270 & 0.5 & 1.42$\cdot$10$^{-3}$\\
 & & 1. & 1.56$\cdot$10$^{-2}$\\
\mr
 & & 0. & 5.2$\cdot$10$^{-4}$\\
$^{166}$Er & 0.287 & 0.5 & 1.59$\cdot$10$^{-3}$\\
 & & 1. & 1.52$\cdot$10$^{-2}$\\
\br
\end{tabular}
\end{indented}
\end{table}
Recalling the fact that at small values of $q$ the slope
of the real Coulomb (magnetic) form-factor allow us to determine the
root-mean-square charge (magnetic) radii [19], we proceed in the
same fashion to seek the transition toroidal quadrupole moment.
According to eq.(23) and the definition of the real transverse electric
form-factor (25) the toroidal quadrupole moment, for an arbitrary value
of the rigidity parameter, reads
\begin{equation}
{\bcal T}_{2}({\tilde r})={3e^{2}Z\over 56\pi}
\frac{(3+2{\tilde r})}{{\bcal I}_{{\tilde r}}}R_{0}^{4}\beta
\end{equation}
Values of this transition moment for different nuclei and
rigidity parameters are given in table 1.
We have also plotted ${\bcal T}_{2}$ versus the rigidity
parameter ${\tilde r}$ for $^{152}$Sm and $^{166}$Er (fig.2). An obvious,
but important conclusion in our view represents the fact that the toroidal
quadrupole moment drastically increase for nuclei which exhibite vortical
components of the flow, being two order in magnitude greater in the RR
case relative to the IF case.

\section{Observation of toroidal quadrupole transitions in
electron scattering}

In view of the definition of toroidal multipoles their measurement is
equivalent to the measurement of the transverse electric multipole at
high-momentum transfer and afterwards the substraction of the Siegert limit.
The use of inelastic electron scattering allows the measurement of form
factors (f.f.) at momentum transfers ranging from the photon
point to $qR\gg 1$. Therefore it is possible to get information about the
transverse f.f. for $q$ values which make the charge-current constraints
important up to $q$-regions where the Siegert theorem breaks up and the
toroidal f.f. dominates the transverse electric part of the cross-section.
The cross-section for inelastic electron scattering on an unpolarized
nucleus can be written in the PWBA as [20]
\begin{eqnarray}
\fl\frac{d\sigma}{d\Omega}=4\pi\sigma_{Mott}{1\over 2I_{i}+1}
\{\frac{q_{\mu}^{4}}{q^{4}}{\sum_{\lambda=0}^{\infty}}
|\langle I_f\mid \mid {\hat M}_{\lambda}(q)\mid \mid
I_i\rangle|^{2}\nonumber\\
\lo +(\frac{q_{\mu}^{2}}{2q^{2}}+\tan^{2}{\theta\over 2})
\sum_{\lambda=1}^{\infty}
|\langle I_f\mid \mid {\hat T_{\lambda}^E}(q)\mid \mid
I_i\rangle|^{2}+
|\langle I_f\mid \mid {\hat T_{\lambda}^M}(q)\mid \mid
I_i\rangle|^{2}\}
\end{eqnarray}
where $\sigma_{Mott}$ is the Mott cross section, $q_{\mu}=({\bi
q},\omega)=
P_{i\mu}-P_{f\mu}$ is the four momentum transfer from the
nucleus, $P_{i\mu}$
and $P_{f\mu}$ are the initial and final nuclear four-momenta,
$q=|{\bi q}|$,
$q_{\mu}^{2}=q^{2}-\omega^{2}$ and ${\hat M}_{\lambda}$,
${\hat T}_{\lambda}^{E}$ and ${\hat T}_{\lambda}^{M}$ are the
Coulomb,
transverse electric and magnetic multipoles respectively. In the case of
exciting the g.s. band of an even-even nucleus there will be involved the
electric multipoles with $\lambda\geq 2$ and the magnetic multipoles with
$\lambda$-odd. In the present work for simplicity we shall discard
the magnetic multipoles, thereby confining ourselves to the analysis of
longitudinal and transverse electric parts of the cross section. For our
present purposes, which consists in the determination of toroidal
transitions in connexion with the possibility of observing nuclear
vortical currents we must chose an appropriate method to subtract the
dominant longitudinal component from the cross section and therefore
obtaining the searched transverse multipoles. The method of separating
the transverse multipole at 180$^{\circ}$ angle scattering is very
seducing, since at such angles they are dominating the cross-section.
During the last times there have been reported measurements of
normal-parity transverse excitations in such experiments [21].
Another method of separating the transverse
multipole from the longitudinal one consists in performing scattering of
polarized electrons on polarized nuclear targets. Such processes allow the
measurement of observables which involve the interferences
between longitudinal and transverse multipoles [3,22]. In this paper we
use the first method because it does not imply the complex calculations
required by the second method.

For the excitation $0^{+}\rightarrow\lambda^{+}$ it turns out the
following expression for the differential cross-section
\begin{equation}
\frac{d\sigma}{d\Omega}=4\pi\sigma_{Mott}f_{rec}^{-1}
\{\frac{q_{\mu}^{4}}{q^{4}}
|{t_{C\lambda}}(q)_{0^{+}\rightarrow\lambda^{+}}|^{2}+
(\frac{q_{\mu}^{2}}{2q^{2}}+\tan^{2}{\theta\over 2})
|{t_{E\lambda}}(q)_{0^{+}\rightarrow\lambda^{+} }|^{2}\}
\end{equation}
Using the multipole decomposition which separates the toroidal multipoles
and neglecting in the above expression of the differential cross-section
the Coulomb multipoles and the low-$q$ limit of the transverse electric
multipole one get
\begin{equation}
\frac{d\sigma}{d\Omega}=4\pi\sigma_{Mott}f_{rec}^{-1}{q^{2}}
({1\over 2}{q_{\mu}^{2}}+{q^{2}}\tan^{2}{\theta\over 2})
|{t_{T\lambda}}(q)_{0^{+}\rightarrow\lambda^{+} }|^{2}
\end{equation}
This approximation is equivalent to neglecting the transition
charge moments $Q_{\lambda}$ and the average 2n-power radii of the charge
distribution $r_{\lambda}^{2n}$ defined in [4,14,19].
We have represented graphically the differential
cross-sections for the RR and IF models in both cases, the exact (29) and
approximate (30), for the quadrupole (fig.3a) and hexadecupole (fig.3b)
transitions induced
by the scattered electron. It is transparent from these figures that the
quoted approximation, works well for the RR model at low and high
momentum transfer for both considered transitions. The principal
difference appears in the location of the diffraction minima which are
lowered in the approximate case relative to the exact one. However this is
not very important since in a phase shift analysis the curve in the
neighborhood of the sharp minimas is smoothed. In the IF case the
agreement between the exact and approximate differential cross-section is
good for $q<400$ MeV/$c$ for $\lambda=2$ and $q<250$ MeV/$c$  for
$\lambda=4$. The reason which determines the discrepancies for the IF
model between the the exact and approximate curves steems on the fact
that at higher momentum transfer the 2n-power radii of the charge
multipoles contribution in the cross-section is enhanced.

\section{Conclusions}

We end-up this paper by substantiating the main results:
\begin{enumerate}
\item The quantity called deviation from the Siegert theorem introduced in
this paper, which describes the contribution of high-$q$ terms in the
expression of the transverse electric multipole contains the information
about the currents inside the nucleus. For the RR which is a submodel
of the more general Riemann model, the deviation
from the Siegert limit has a larger slope than the corresponding quantity
defined for the IF model. The deviation from the Siegert
theorem increase with the rigidity parameter ${\tilde r}$ and the mass
number A for the considered nuclei in the range $q=0\div 150$ MeV/$c$
and oscillates around $\eta=1$ above $q=150$ MeV/$c$.
\item The transition toroidal quadrupole moment is a monotous increasing
function of the vorticity. This strengths our opinion that the toroidal
moments are to be associated with vortical electromagnetic currents in the
same way as the charge multipole moments are associated with charge
distribution or irrotational electromagnetic currents.
\item At backward electron scattering the differential cross section of the
RR model may be approximated by taking into account only the toroidal
form factor for a broad range on momentum transfer.
\end{enumerate}

\ack I would like to express my gratitude to A.A. R\u{a}du\c t\u{a} for
reading the manuscript and his interesting remarks.
I am also expressing my thanks for valuable discussions with V.M. Dubovik,
P.G.H. Sandars, J.D. Walecka, E. Moya de Guerra, N. Lo Judice and
R. Liotta. The assistance of A.Andronic is acknowledged.

\Figures
\Figure{The deviation from the Siegert theorem for the the
nucleus
$^{166}$Er for the excitations $0^{+}\rightarrow 2^{+}$
($\lambda=2$) and
$0^{+}\rightarrow 4^{+}$ ($\lambda=4$) considering the case of IF
and RR.}
\Figure{Transition toroidal quadrupole moment dependence on the
rigidity
parameter for $^{152}$Sm($\full$) and $^{166}$Er($\broken$)}.
\Figure{Differential cross section versus momentum transfer for
(a) $\lambda=2$ and (b) $\lambda=4$ for both IF and RR. For each
model two
sets of datta are ploted. One is obtained by using eq.(29) which
includes the
Coulomb ans transverse electric form-factors the other use only
the
toroidal form factor (eq.(30)).}

\end{document}